\documentclass[prl,twocolumn]{revtex4-1}

\pdfoutput=1 
\usepackage{graphics} 
\usepackage{graphicx} 

\begin{document} 

\title{Computing Gibbs free energy differences by interface pinning}
\email{ulf.pedersen@tuwien.ac.at} 

\author{Ulf R. Pedersen$^{1,2}$} 
\author{Felix Hummel$^{2}$} 
\author{Georg Kresse$^{2}$} 
\author{Gerhard Kahl$^{1}$} 
\author{Christoph Dellago$^{2}$} 
\affiliation{$^1$Institute of Theoretical Physics, Vienna University of Technology, Wiedner Hauptstra{\ss}e 8-10, A-1040 Vienna, Austria}
\affiliation{$^2$Faculty of Physics, University of Vienna and Center for Computational Materials Science, Sensengasse 8/12, A-1090 Vienna, Austria}

\date{\today} 

\begin{abstract}
We propose an approach for computing the
Gibbs free energy difference between phases of a
material. The method is based on the determination of the average
force acting on interfaces that separate the two phases of
interest. This force, which depends on the Gibbs free energy
difference between the phases, is computed by applying an external
harmonic field that couples to a parameter which specifies the two
phases. Validated first for the Lennard-Jones model, we demonstrate the
flexibility, efficiency and practical applicability of this approach
by computing the melting temperatures of sodium, magnesium, aluminum and silicon
 at ambient pressure  using density functional theory. Excellent agreement with experiment is found for 
all four elements, except for silicon, for which the melting temperature is, 
in agreement with previous simulations, seriously underestimated.
\end{abstract}

\maketitle

An accurate location of first order transition lines at a reasonable
computational cost is of paramount importance for a wide spectrum of
condensed matter systems, ranging from hard to soft materials and
biological matter.  Basic principles of equilibrium thermodynamics
imply that for a given temperature and pressure the system resides in
the phase of lowest Gibbs free energy. Phase transitions occur where
Gibbs free energy differences between phases vanish, determining phase
boundaries in the pressure-temperature plane. From the computational
point of view, however, the task of evaluating a phase diagram
represents a significant challenge, as phase transitions occur on long
time scales \cite{becker1935} such that they cannot be studied using
straightforward molecular dynamics or Monte Carlo simulations.

Several numerical approaches have been
proposed to cope with this problem \cite{frenkel2002,vega2008}: (i) in
the {\it indirect approach}, often based on thermodynamic integration,
the Gibbs free energy is computed
individually for each of the phases
\cite{widom1963,hoover1968,hoover1971,norman1996,alfe2003} and the
coexistence line is then calculated by imposing the coexistence
condition of equal Gibbs free energy. (ii) Alternatively, in the {\it
  direct approach}, an explicit interface is introduced between the
two phases which are then simulated simultaneously in the same
simulation box. At fixed pressure and temperature, the system moves towards the phase
with the lower Gibbs free energy. Exactly at coexistence the thermodynamic
driving force on the interface vanishes and the interface stops moving 
except for thermal fluctuations. Successful applications of this
approach have been reported for a broad spectrum of materials
\cite{ladd1977,broughton1986,landman1986,mori1995,kyrlidis1995,agrawal2003,morris2002,hoyt2002,sibugaga2002,fernandez2006,weingarten2009,timan2010,pedersen2011_lwotp}.

In this contribution, we present and validate a method to
compute the Gibbs free energy difference, $\Delta G$, between two
phases. The basic idea of this approach is to compute the average
force required to {\it pin} the interface of a two-phase system via a
harmonic bias potential. 
This external field couples to a suitably defined order parameter, $Q$, 
which distinguishes between the phases of interest. The application of the bias 
potential effectively transforms the out-of-equilibrium process of the 
conventional moving interface method into a well-defined equilibrium computation, 
in which the free energy difference $\Delta G$ is determined directly. 
We refer to this approach as the ``interface pinning'' method. Coexistence 
points may subsequently be determined using Newton's root finding method.

To validate our new approach, we have first applied it to the Lennard-Jones (LJ) model\cite{lennard-jones1924}. Our calculations
reproduce with high accuracy the solid-liquid coexistence line identified previously with other approaches
\cite{hansen1969,morris2002,mastny2007} and provide Gibbs free energies that are in excellent agreement 
with those obtained by thermodynamic integration. We have then used interface pinning in 
combination with {\it ab initio} simulations to compute the melting temperatures of sodium (Na), magnesium (Mg), aluminum (Al) and silicon (Si), demonstrating that this method is efficient, flexible and widely applicable.

Compared to the conventional direct and indirect methods used in the literature so far, interface 
pinning offers several advantages. In contrast to the direct approaches, interface pinning 
operates at well-defined equilibrium conditions, thus permitting the explicit calculation of free energy differences
and interface properties.
The selection of the order parameter $Q$ does not need to be a reaction coordinate capturing the entire transformation mechanism. 
Finally, interface pinning inherits the general applicability and conceptual simplicity of the direct approaches. The latter makes it easy to implementation into existing programs.

\begin{figure} 
\begin{center} 

  \includegraphics[width=0.5\columnwidth]{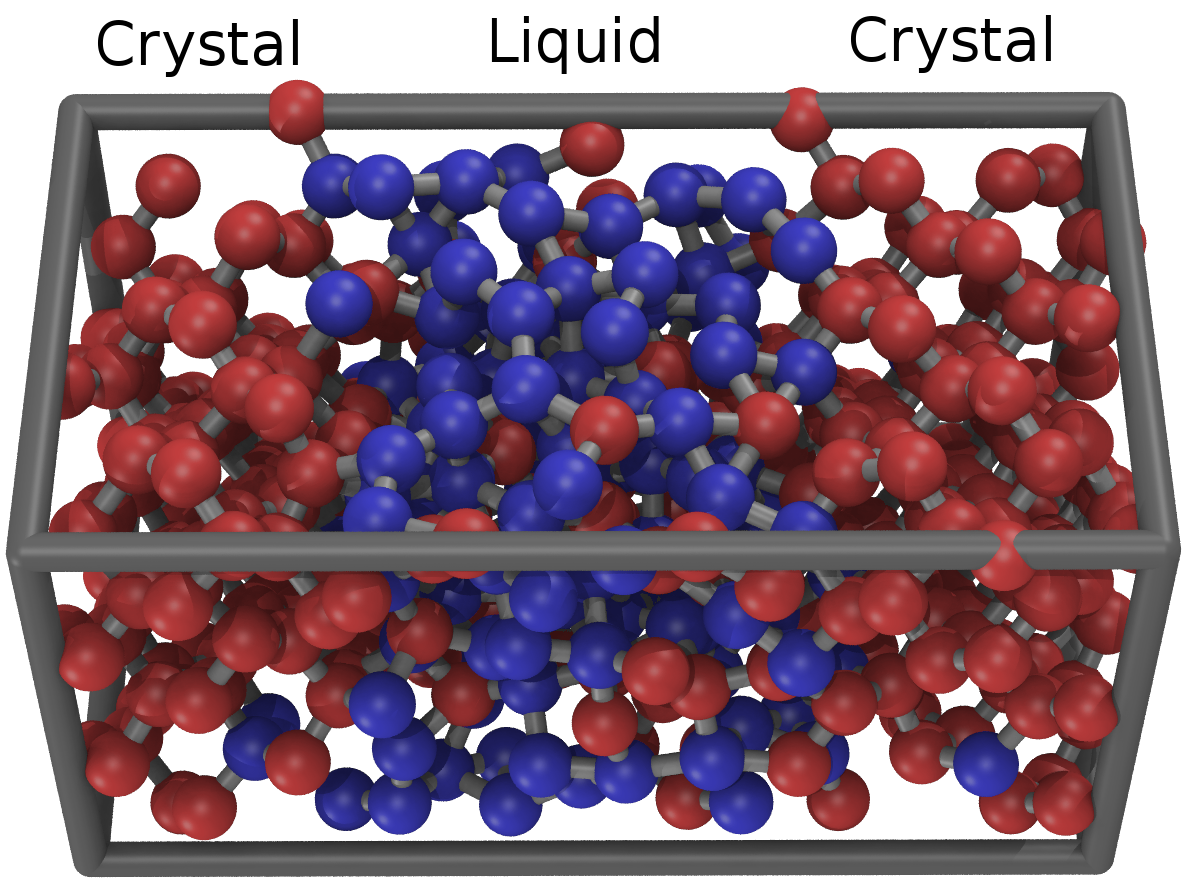} 
  \includegraphics[width=0.75\columnwidth]{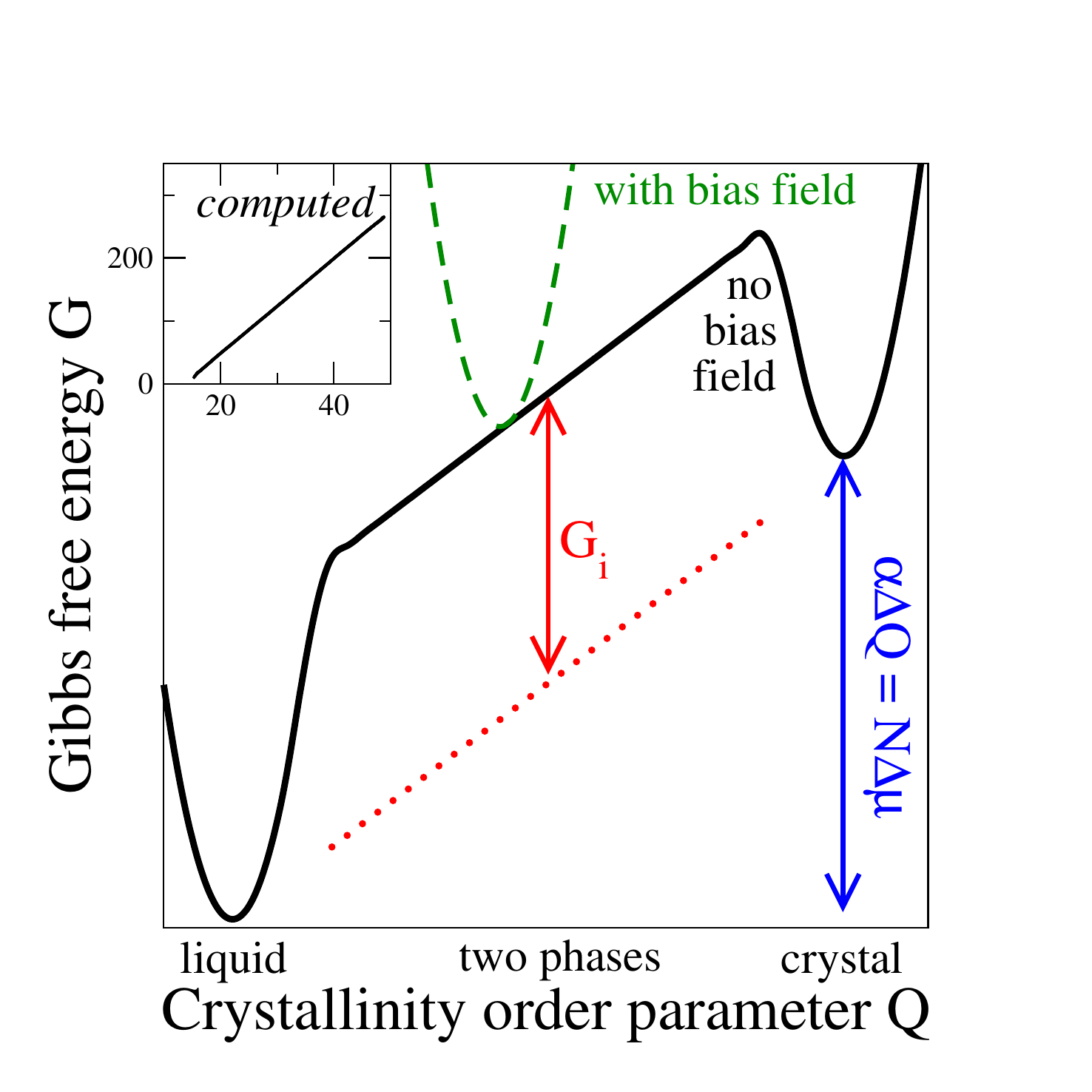} 

  \caption{\label{sketch} (color online). Upper panel: crystal-liquid
    configuration from an {\it ab initio} simulation of 432 Si atoms
    in a periodic box. Atoms are colored according to the coordination number (red=[fourfold coordinated] and blue otherwise).
    Lower panel:
    schematic sketch of the Gibbs free energy $G(Q)$ (black solid line)
    as a function of the crystallinity order parameter $Q$ at a
    state point where the liquid is thermodynamically stable and the
    crystal is metastable. The double arrows indicate the interface
    contribution $G_i$ (red) and the bulk contribution $N\Delta \mu$
    (blue), respectively. The dashed green curve indicates the Gibbs
    free energy $G'(Q)$ with bias potential
    applied. The inset shows that the computed
    $G(Q)/ (k_B T)$ in the two-phase region is indeed linear; here, $G(Q)$ was computed for the LJ model ($N=5120$) via umbrella sampling \cite{frenkel2002,shirt2008} using
    Equ. (\ref{EnergyWithField}) with $\kappa=2$ and a range of $a$'s.}
\end{center} 
\end{figure}

To introduce the method, consider a two phase crystal-liquid system
\cite{woodruff1973} in a periodic orthorhombic box (see figure
\ref{sketch}) at temperature $T$ and pressure $p$. The box lengths $X$
and $Y$ are kept constant at values for which the crystal is
unstrained, while the box length $Z$ is allowed to change in order to
maintain constant pressure.  We refer to this ensemble as the
$Np_zT$-ensemble. To lower the interface Gibbs free energy $G_i$,
the system will have two interfaces in the $XY$-plane minimizing the
interface surface area.  We assume that the system is large enough to
represent bulk phases at least at the center of the liquid and solid slabs.
Particles may then either be labeled
crystalline (subscript $c$), liquid (subscript $l$) or interfacial
(subscript $i$), so that the total number of particles is $N = N_c +
N_l + N_i$.  The contributions to the total Gibbs free energy of
particles in the bulk phases is determined by the chemical potentials
$\mu_c$ and $\mu_l$ of the solid and liquid, respectively, and the
total Gibbs free energy is $G = N_c\mu_c + N_l\mu_l + G_i$.

When the relative distance
between the interfaces changes, particles are transferred between the
bulk phases. Assuming that the interface quantities $G_i$ and $N_i$ do not change when the 
interfaces shift due to the growth of one bulk phase at the cost of the other, the number of 
liquid particles may be written as $N_l=-N_c+\textrm{[constant]}$ and the Gibbs free energy 
is given by
\begin{equation}\label{GofN}
G(N_c)=N_c \Delta \mu + \textrm{[constant]}
\end{equation}
where $\Delta \mu \equiv \mu_c-\mu_l$. Throughout the paper we will
use the subscripts $c$ and $l$ for crystal and liquid properties,
respectively, and let ``$\Delta$'' denote
``$\textrm{[crystal]}-\textrm{[liquid]}$''.

To sample configurations in the two-phase region and to prevent the
system from complete transformation into one of the pure phases, we
apply a harmonic bias potential that pins the relative position of the
interfaces. Let $U({\bf R})$ be the energy of the unbiased system for 
configuration ${\bf R}=\{{\bf r}_1,{\bf r}_2,\ldots,{\bf r}_N\}$, and
\begin{equation}
  U'({\bf R})=U({\bf R})+\frac{\kappa}{2}[Q({\bf R})-a]^2,
  \label{EnergyWithField}
\end{equation}
the energy of the system plus the bias potential. Here, $Q({\bf R})$ is a
global order parameter with a linear dependence on the number of particles in the solid phase: $Q=\frac{N_c}{N}Q_c+\frac{N_l}{N}Q_l+\frac{N_i}{N}Q_i$ so that
\begin{equation}
N_c=N\frac{Q}{\Delta Q}+\textrm{[constant]}.
  \label{NQlinear}
\end{equation}
In the biased system, the position of
the interfaces relative to each other will fluctuate around an average value and the order
parameter $Q$ will fluctuate accordingly. The probability distribution
of $Q$ is $P'(Q)=\exp[-G'(Q)/k_BT]/{\cal Z}'$ where $G'(Q)$ is the
Gibbs free energy along the $Q$ coordinate of the biased system, and
${\cal Z}'$ is the corresponding partition function. The Gibbs free
energy of the biased system may be written in terms of the unbiased
free energy as $G'(Q)=G(Q)+\frac{\kappa}{2}(Q-a)^2+k_BT\ln({\cal
  Z}'/{\cal Z})$. By insertion of Equs. (\ref{GofN}) and (\ref{NQlinear}), it follows
that $P'(Q)$ is Gaussian,
\begin{equation}
P'(Q) = \sqrt{\frac{k_BT}{2\kappa\pi}} \exp\left\{-\frac{\kappa}{2k_BT}[Q-a+\alpha/\kappa]^2\right\},
\end{equation}
where $\alpha=N\Delta\mu/\Delta Q$ is the slope of $G(Q)$ in the
two-phase region, displayed in the lower panel of Fig. \ref{sketch}.
The distribution $P'(Q)$ has variance $\sigma^2_Q=k_BT/\kappa$ and average $\langle Q\rangle'=a-\alpha/\kappa$, and the
chemical potential difference between the two phases may be computed
as
\begin{equation}\label{dmu} 
\Delta\mu = -\kappa(\langle Q\rangle'-a)\Delta Q/N. 
\end{equation}
As a guideline, we choose $\kappa$ such that typical fluctuations in 
$Q$ correspond to one or a fraction of a crystal plane, and $a$ such 
that the system is approximately half liquid and half crystal. In 
practice, we find that a wide range of field parameters give the same 
precision of the $\Delta \mu$ estimate \cite{pedersen_long}. 

Once $\Delta\mu$ is known, coexistence points may be
determined using Newton's method for finding roots. The required
derivatives of $\Delta\mu$ along isobars and isotherms
is given by the standard thermodynamic expressions,
$\partial(\Delta\mu)/\partial p|_T=\Delta v$ and
$\partial(\Delta\mu)/\partial T|_p=-\Delta s=-(\Delta u + p
\Delta v - \Delta\mu)/T$. In these relations, $\Delta v$, $\Delta s$, and $\Delta u$ are changes in specific volume,
entropy, and energy, respectively.

To apply the interface pinning method in practice, we must choose an
order parameter $Q$ that grows linearly with the number of crystalline
particles $N_c$ in the two-phase region. Moreover, $Q$ should be
computationally inexpensive. Unlike liquids, crystals have long-ranged
translational order, allowing us to use the collective density field
as order parameter: $Q=|\rho_{\bf k}|$ where $\rho_{\bf k} = N^{-\frac{1}{2}} \sum_{j=1}^N \exp ( -i{\bf k}\cdot{\bf r}_j )$.
%
%
Here, ${\bf k}=(2\pi n_x/X,2\pi n_y/Y,0)$ for some fixed integers
$(n_x,n_y)$ that should be chosen such that ${\bf k}$ correspond to a
Bragg peak. This choice will maximize the contrast between the liquid
and the crystal. The $z$-component of ${\bf k}$ is set to zero since
$Z$ fluctuates in the $Np_zT$-ensemble.  
The constant $N^{-\frac{1}{2}}$ makes $Q_l$ system size
invariant while $Q_c\propto N^\frac{1}{2}$. Derivatives of $Q$ with respect to 
the particle coordinates, required to determine the forces resulting from 
the bias, can be computed with an algorithm scaling as $O(N)$. 
We note that this order parameter may be problematic in the supercooled regime, since a crystal can lower $|\rho_\mathbf{k}|$ by introducing long wave length displacements of particles. The energy penalty of such displacements is low and decreases with increasing system size. We have chosen to use $|\rho_{\bf k}|$ as order parameter for most computations, since it is generally applicable and simple. For some computations we have in addition used the Steinhardt $Q=Q_6$ order parameter \cite{steinhardt1983}, which has the advantages of being robust in the supercooled regime. The two choices of order parameter give the same $\Delta\mu$'s within statistical error. A more detailed description of the method will be given in a forthcoming publication \cite{pedersen_long}.

\begin{figure} 
\begin{center} 
  \includegraphics[width=0.75\columnwidth]{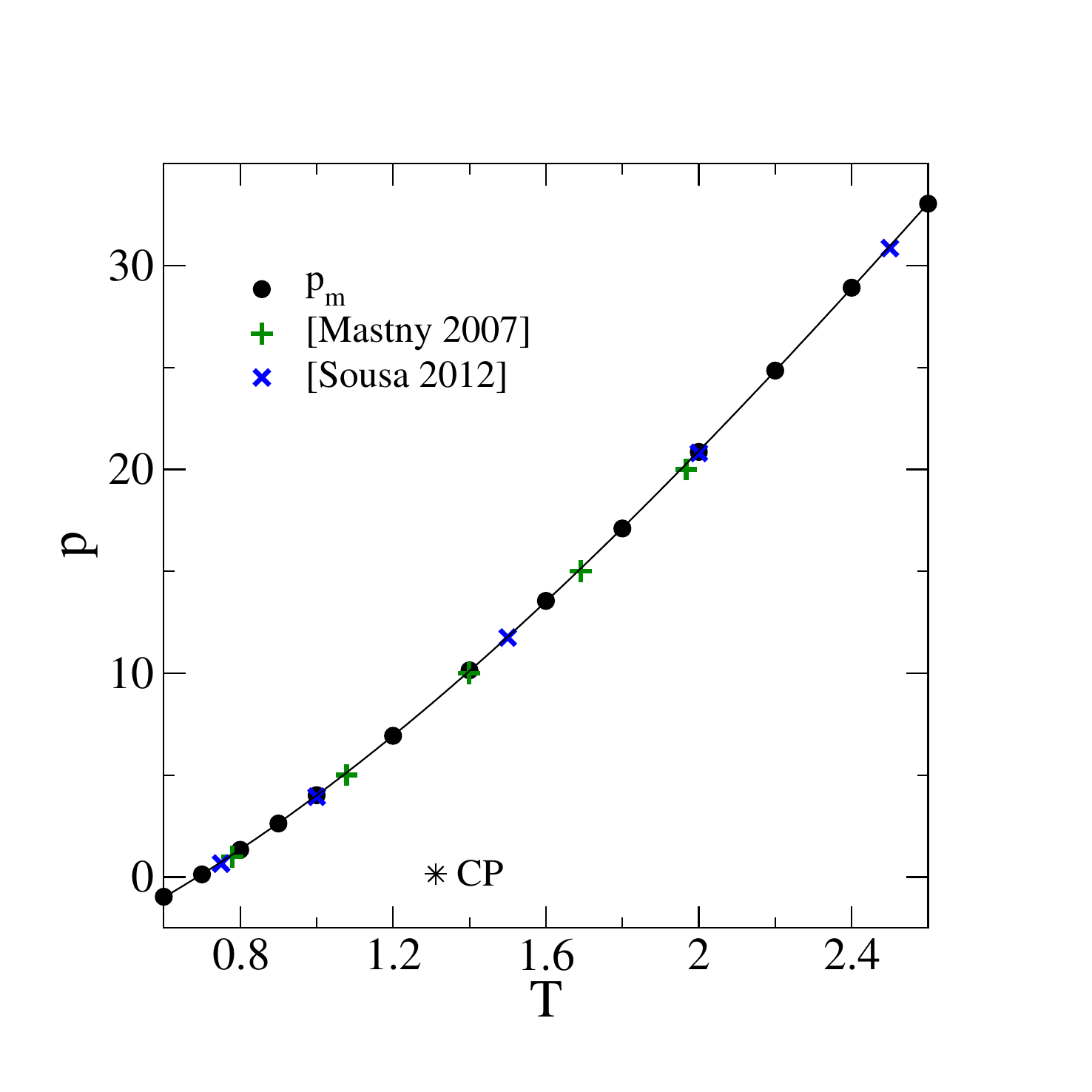}
  \caption{\label{coex_line} (color online). Crystal-liquid coexistence of the LJ model (filled black dots) in the $(p,
    T)$-plane computed with interface pinning for the LJ model. The solid line
    is a cubic fit: $-0.5223T^3+5.017T^2+5.502T-5.989$. The computed coexistence line agrees well with results of other methods \cite{p_long}: $+$'s and $\times$'s are from Refs.~\cite{mastny2007} and \cite{sousa2012}, respectively. The asterisk indicates the gas-liquid critical point ($T_\textrm{CP}=1.31$; $p_\textrm{CP}=0.15$) of the full LJ model \cite{potoff1998}.
    }
\end{center} 
\end{figure} 

\begin{figure} 
\begin{center} 
  \includegraphics[width=0.75\columnwidth]{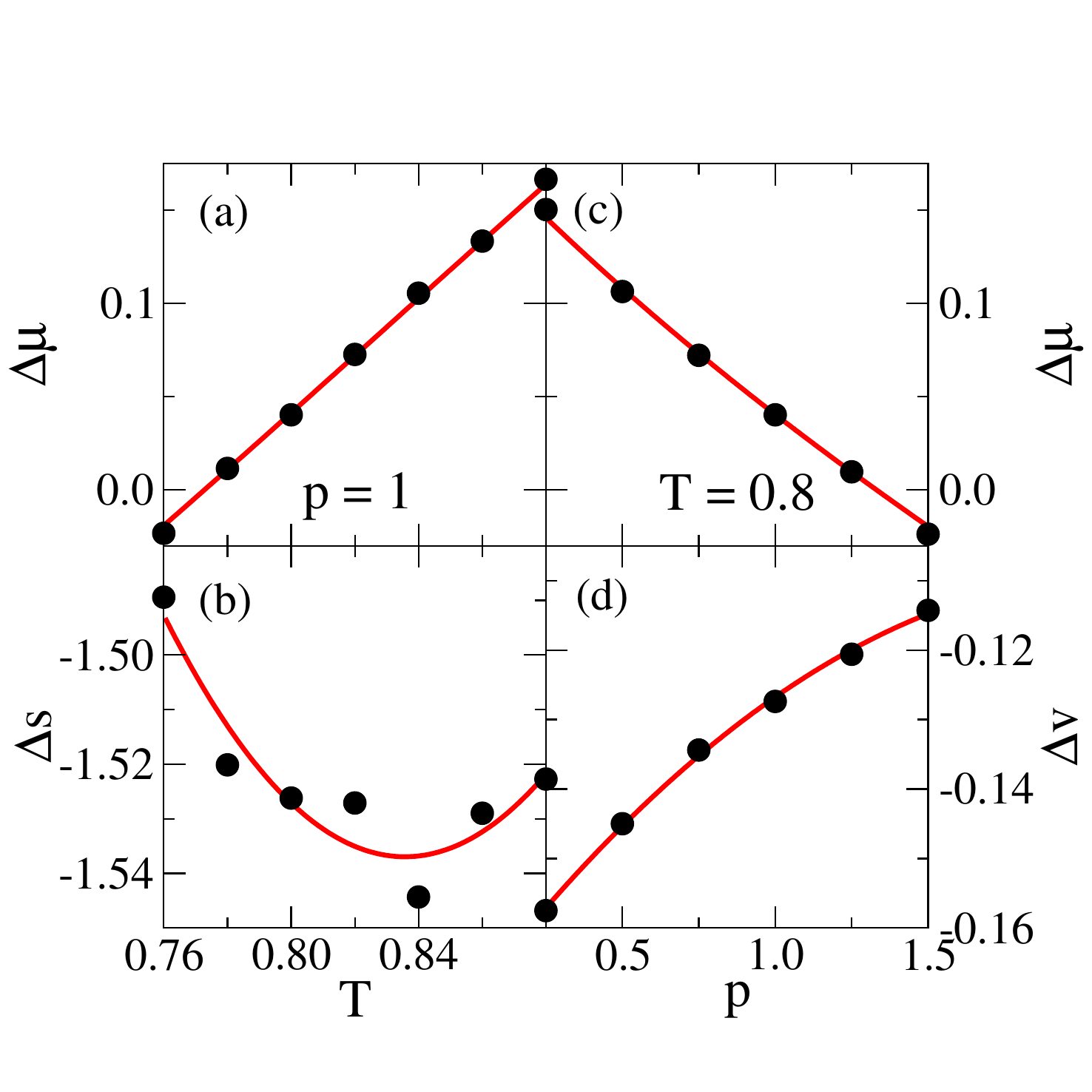}
  \caption{\label{consist} 
    (color online). Upper panels (a and c): $\Delta\mu$ computed with
    interface pinning method along an isobar and an isotherm,
    respectively. Lower panels (b and d): 
    specific entropy $\Delta s$ vs. $T$ and specific volume $\Delta v$ vs. $p$,
    respectively; the solid lines in the lower panels are quadratic
    polynomial fits to these data. The solid lines in the upper
    panels were computed by integration of these fits. The
    integration constants were chosen to provide the best overall
    agreement with the $\Delta \mu$-data.}
\end{center} 
\end{figure}

To verify the method, we first used it to determine the solid-liquid coexistence line of the LJ model with truncated pair interactions: $U({\bf R}) =
\sum^N_{i>j}u(r_{ij})$, where $u(r) =
4(r^{-12}-r^{-6})-4(6^{-12}-6^{-6})$ for $r<6$
and zero otherwise (LJ units are used for this model throughout the
paper). MD simulations with a time step of $t_{\rm step} = 0.004$ were performed using the LAMMPS software package
\cite{lammps} modified to include the bias potential.  
The Parrinello-Rahman barostat was used \cite{parrinello81}
with a time constant of $\tau_{\rm PR}=8$ together with a
Nos{\'e}-Hoover \cite{nose1984,hoover1985} thermostat with a time constant of $\tau_{\rm NH}=4$.

Results presented in Fig. \ref{coex_line} demonstrate that the solid-liquid coexistence line for
the LJ model computed by interface pinning agrees within high
precision with data obtained using other methods
\cite{mastny2007,sousa2012}. The coexistence points displayed in Fig. \ref{coex_line} were
computed as follows. First, a crystal structure of 8$\times$8$\times$20
face centered cubic (fcc) unit cells ($N=5120$) was constructed and
simulated at $p=1$ and $T=0.8$ for $t_{\rm sim}=800$. All box lengths were allowed to fluctuate in order to
determine the geometry of the unstrained crystal, giving $X=Y=12.85$.
The unstrained crystal was then simulated for
$t_{\rm sim}=800$ in the $Np_zT$ ensemble, and $Q_c=56.31$ $(n_x=16,n_y=0)$ and the average partial volume $v_c=1.036$ was
recorded. Next, a liquid was prepared by melting the
crystal in a constant volume simulation at high temperature ($T=5$). The
$Np_zT$-ensemble (using $X=Y=12.85$) of the liquid was simulated for
$t_{\rm sim}=800$, and $Q_l=0.94$ and the average specific
volume of the liquid $v_l=1.163$ was recorded. Then, a two phase
configuration was constructed by performing a high temperature
constant volume simulation where particles at $z<Z/2$ were kept at
their crystal positions using harmonic springs anchored at crystal
sites, with the box volume (length $Z$) in
between that of the crystal and the liquid. The $Np_zT$-ensemble
with the bias-field of Equ. (\ref{EnergyWithField}) with parameters $a=26$ and
$\kappa=4$ was simulated for $t_{\rm sim}=4000$ to compute
$\langle Q\rangle'= 25.055$. Application of Equ. (\ref{dmu}) yielded a chemical
potential difference of $\Delta\mu = 0.040$.
The coexistence pressure was then determined iteratively using Newton's root 
finding method along the isotherm: $p^{(i+1)} =
p^{(i)} - \Delta\mu^{(i)}/\Delta v^{(i)} $, providing pressures of 
$p^{(i)} =\{1.0, 1.320, 1.337(1)\}$.  In the last iteration the estimated chemical
potential difference is zero within the error bar,
$\Delta\mu=-0.0007(10)$ (throughout the paper numbers in parentheses indicate the statistical errors of the last digits). 
In Fig. \ref{consist} we confirm that $\Delta\mu(p,T)$ computed with
interface pinning (symbols) is consistent with thermodynamic integration (lines).

\begin{table} 
\caption{\label{tbl_ab_initio}{\it{Ab inito}} and experimental (Refs. \onlinecite{hultgren1973,chase1985}) 
melting temperatures $T_m$ (in K) of
period 3 elements using either $|\rho_k|$ or $Q_6$ as order parameter.
"Super cell" indicates the applied super cell built from 
the conventional cubic cell (including liquid and solid part). $N$ is the total particle number.}
\begin{ruledtabular} 
\begin{tabular}{ r | c c c c c c  } 
 & unit cell & super cell   & $N$ & $Q$  & $T_m$ [exp.] \\
\hline
Na & bcc    & 5$\times$5$\times$10 & 500 &$|\rho_{\bf k}|$ \& $Q_6$ &   354(21) [370]  \\
Mg & hcp & 4$\times$6$\times$8$^1$ & 767 &$|\rho_{\bf k}|$ \& $Q_6$ &   920(20) [923]   \\
Al & fcc   & 4$\times$4$\times$8  & 512 &$Q_6$&  985(30) [933] \\
Si & cd    & 3$\times$4$\times$7 & 672   &$|\rho_{\bf k}|$&  1241(20) [1635] \\
\end{tabular}
\end{ruledtabular}
\\ $^1$ built from an orthorhombic 4 atom $(a, \sqrt{3} a, c)$ cell.  
\end{table}

Due to its efficiency and flexibility, the interface pinning approach can be combined 
with electronic structure methods and {\it ab initio} molecular dynamics to computed free energy differences from first principles.
For the present simulations, the method was implemented in the Vienna {\it ab inito} Simulation Package \cite{kresse1996}. 
As an example, we used interface pinning to compute the melting temperatures $T_m$ of the period three elements Na, Mg, Al and Si at ambient pressure. Computed $T_m$'s are shown in Table \ref{tbl_ab_initio} along with simulation details. Melting temperatures were computed first for crystalline Si in the fourfold coordinated cubic diamond (cd) structure (see Fig. \ref{sketch}). 
To be compatible to previous calculations \cite{sugino1995,alfe2003}, density functional theory (DFT) in the local 
density approximation (LDA) within the framework of the projector augmented wave method was used \cite{bloechl1994}. $NpT$ and $Np_zT$ simulations were performed using a time step of
$t_{\rm step} =3$ fs with a Parrinello-Rahman barostat \cite{parrinello81}
and a Langevin thermostat \cite{langevin1908}. 
To compute the Si coexistence temperature at ambient pressure, we use a similar strategy as outlined for the LJ model: bulk properties of the crystal and the liquid ($Q_c$, $Q_l$, $v_c$, $v_l$, $X$ and $Y$) were evaluated in simulations for 216 Si atoms (3$\times$3$\times$3 conventional cells; $t_{\rm sim}=60$\,ps) at $T=1200$ K. Next, solid-liquid simulations with a bias field ($t_{\rm sim}>30$\,ps) were performed for four system sizes: \{2$\times$2$\times$4, 2$\times$2$\times$7, 3$\times$3$\times$6, 3$\times$4$\times$7\} conventional cubic cells  corresponding to $N=\{128,224,432,672\}$ atoms. Coexistence temperatures were estimated to be \{1189,1218,1225,1241\} K using $T_m\simeq T +\frac{\Delta \mu}{\Delta s}$. Finally, finite size effects were extrapolated assuming a  $1/N$ decay of the finite size error  yielding $T_m=1250(10)$\,K. The finite
size effects are particularly large for liquid silicon, since the metallic liquid is
embedded in a semiconducting host, resulting in a discretization of the electronic states in the metal
(electron in a box).
The present value is fully consistent with previous LDA calculations~\cite{sugino1995,alfe2003}, 
and the  discrepancy  to the
experimental value of $T_m=1635\,$K originates from an
underestimation of the energy difference between four fold coordinated semiconducting Si in the cd structure
and  six fold coordinated metallic Si in liquid Si resembling  the  $\beta$-tin structure \cite{alfe2003}. 
The entropy of fusion $\Delta s(T_m) = 3.5(1)\,k_B$/atom and the slope of the melting curve $d\,T_{\rm m}/d\,p = -51(7)\,$K/GPa
(computed using the Clausius-Clapeyron relation) are also in
agreement with previous theoretical results \cite{sugino1995,alfe2003}.
For the other elements, Na, Mg and Al, finite size effects are less critical, and we only considered  system sizes comparable to the largest Si system. For these three elements, the calculations were performed using PBEsol (Perdew, Burke, Ernzerhof functional for solids)\cite{PBEsol}, which
yields more accurate lattice constants than the LDA.
The computed $T_m$'s of Na and Mg are in excellent agreement with experimental values, while for Al the computed $T_m$ is about 6\% too large  (see Table \ref{tbl_ab_initio}).

In summary, we have introduced a computational method that allows a direct
evaluation of the Gibbs free energy differences between two phases. In
contrast to previous approaches, simulations are carried out
at equilibrium conditions by pinning the interface between the phases of interest via a harmonic
bias potential that couples to a suitably defined order
parameter. Application of interface pinning to the LJ model demonstrates the accuracy and
efficiency of this new approach: the solid-liquid coexistence line 
agrees to high accuracy with data obtained by other methods and the 
computed Gibbs free energy is consistent with data obtained via 
thermodynamic integration. The practical applicability and 
flexibility of the method was demonstrated by computing the 
melting points of Na, Mg, Al and Si at ambient pressure using first principles simulations.  The results demonstrate
that present density functionals yield very accurate melting temperatures for crystalline metal to liquid metal
transitions, but errors are sizable for semiconductor (cd-Si) to metal transitions (liquid Si). 
Furthermore, the present approach allows to compute directly and straightforwardly  
structural and thermodynamic properties of the interface, such as surface tension from the pressure tensor \cite{kirkwood1949} or the crystal growth rate from $Q(t)$ fluctuations \cite{briels1997,tepper2002}.

This work was financially supported by the Austrian Science Fund FWF
within the SFB ViCoM (F41). Supercomputing time on the
Vienna Scientific cluster (VSC) is gratefully acknowledged.

\bibliography{references}

\begin{thebibliography}{10}
\bibitem{p_long}%
  \BibitemOpen
  \bibinfo {note} {We notice small systematic deviations of $\Delta p_m<0.1$
  related to truncation of LJ interactions.}%
  \bibAnnoteFile{Stop}{p_long}%
\end{thebibliography}

\end{document}